\documentclass[12pt,a4paper,final]{iopart}
\usepackage{iopams}  
\usepackage{graphicx}
\usepackage{hyperref}
\usepackage{cleveref}
\begin{document}
\title[Einstein Flow]{Asymptotic behavior of a matter filled universe with exotic topology}
\author{Puskar Mondal$^{1}$}

\begin{abstract}
The ADM formalism together with a constant mean curvature (CMC) temporal gauge is used to derive the monotonic decay of a weak Lyapunov function of the Einstein dynamical equations in an expanding universe with a positive cosmological constant and matter sources satisfying suitable energy conditions. While such a Lyapunov function does not, in general, represent a true Hamiltonian of the matter-coupled gravity dynamics (unlike in the vacuum case when it does), it can nevertheless be used to study the asymptotic behavior of the spacetimes. The Lyapunov function attains its infimum only in the limit that the matter sources be `turned off` or, at least, become asymptotically negligible provided that the universe model does not re-collapse and form singularities. Later we specialize our result to the case of a perfect fluid which satisfies the desired energy conditions. However, even in this special case, we show using Shutz's velocity potential formalism cast into Hamiltonian form that unlike the vacuum spacetimes (with or without a positive cosmological constant), construction of a true Hamiltonian for the dynamics in constant mean curvature temporal gauge is difficult and therefore the Lyapunov function does not have a straightforward physical interpretation. Nevertheless, we show, for the fluid with equation of state $P=(\gamma-1)\rho$ ($1\leq\gamma\leq2$), that the general results obtained hold true and the infimum of the weak Lyapunov function can be related to the Sigma constant, a topological invariant of the manifold. 
Utilizing these results, some general conclusions are drawn regarding the asymptotic state of the universe and the dynamical control of the allowed spatial topologies in the cosmological models.  
\end{abstract}

Construction of a complete cosmological model of the universe entails the study of the full Einstein equations with suitable matter sources. The so-called FLRW models are rather very special examples of cosmological models, where one takes the cosmological principle literally that the universe appears spatially isotropic and homogeneous if viewed on a sufficiently coarse-grained scale \cite {beesham1994bianchi, nussbaumer2009discovering, ellis1971relativistic}. However, the astronomical observations that motivate the cosmological principle are necessarily limited to a fraction (possibly small) of the entire universe and such observations are compatible with spatial metrics being locally but not globally homogeneous and isotropic. Once the restriction on the `global' topology is removed, there are compact variants of all of the basic FLRW cosmological models, mathematically constructible by taking suitable compact quotients of the basic topologies i.e., Euclidean 3-space $\mathbf{E}^{3}$, Hyperbolic 3-space $\mathbf{H}^{3}$, and 3-sphere $\mathbf{S}^{3}$. A natural question that then arises is whether the governing principle of general relativity does indeed support such cosmological models allowing `exotic' topologies. Recent studies \cite {ashtekar2015general, moncrief2019could} uncovered a surprising dynamical mechanism at work within the framework of the vacuum Einstein flow (with and without a positive cosmological constant $\Lambda$) that strongly suggests that many closed 3-manifolds that do not admit a locally homogeneous and isotropic metric at all will nevertheless evolve in such a way as to be asymptotically compatible with the observed homogeneity and isotropy of the universe. Such a striking result naturally necessitates the inclusion of suitable matter sources in the Einsteinian evolution and the deduction of whether such a matter-filled universe does exhibit the asymptotic behavior compatible with the cosmological observations. Numerous matter models such as dust \cite {buchert2000average}, perfect fluid \cite {ray1972lagrangian}, and Vlasov type matter \cite {ringstrom2013topology}, have been suggested for the cosmological models. In this study, we will consider matter models in the ADM Einstein equations satisfying suitable energy conditions. 

The ADM formulation of the Einstein equations, developed by Arnowitt, Deser, and Misner \cite {arnowitt2008republication}, was originally restricted to vacuum space-times. Taub \cite {taub1954general} developed a gauge fixed Lagrangian variational principle for the relativistic perfect fluid. Later Schutz \cite {schutz1970perfect} used velocity potential formulation to construct a gauge free variational formalism which was still Lagrangian. Afterwards Moncrief \cite {moncrief1977hamiltonian} and Moncrief $\&$ Demaret\cite {demaret1980hamiltonian} developed the Hamiltonian formalism for a perfect fluid and Moncrief $\&$ Demaret \cite {demaret1980hamiltonian} used Taub's co-moving gauge to derive a reduced Hamiltonian (a true Hamiltonian of the dynamics) and it was applied for canonical quantization of general relativistic configurations filled with perfect fluid using Dirac's technique \cite {dirac2001lectures}. However, these studies do not address the topology and the future stability of the universe filled with perfect fluid, and, in particular, do not focus on constructing a suitable cosmological model. On the other hand, \cite {fischer2002hamiltonian, ashtekar2015general, moncrief2019could} constructed a reduced Hamiltonian (respecting the topological restriction) and showed its monotonic decay in the expanding direction in a suitably chosen constant mean curvature gauge. The reduced Hamiltonian (also a weak Lyapunov function), in these studies, seeks to attain its infimum and can achieve its infimum asymptotically, yielding spacetimes admitting spatial slices locally compatible with the cosmological principle. However, such studies are devoid of matter sources i.e., only vacuum spacetimes (with and without cosmological constant) are considered.    

It, therefore, seems necessary to focus on the construction of a suitable Lyapunov function and study its temporal behavior in a general framework which includes a matter source along with a positive cosmological constant. We start with the ADM formalism including a general matter source and later fix a temporal gauge namely the constant mean curvature gauge. Thereafter we study several aspects of dynamical behavior associated with matter sources and assert that the obtained results hold provided that the matter satisfies physically `reasonable' energy conditions. Later on, we specialize our results to a perfect fluid source (which satisfies the required energy conditions) and obtain the asymptotic solutions for which the Lyapunov function stays constant in the limit where the baryon density becomes negligible. Finally, we relate the asymptotic behavior of this Lyapunov function to a topological property of the manifold thereby drawing some general conclusions regarding possible `exotic' topologies of the physical `3+1' universe compatible asymptotically with the cosmological principle.   

\section{ADM formalism with  a Matter source}         
The ADM formalism splits the spacetime described by an `n+1' dimensional Lorentzian manifold $\tilde{M}$ into $\mathbf{R}\times M$ with each level set $\{t\}\times M$ of the time function $t$ being an orientable n-manifold diffeomorphic to a Cauchy hypersurface (assuming the spacetime admits a Cauchy hypersurface) and equipped with a Riemannian metric. Such a split may be executed by introducing a lapse function $N$ and shift vector field $X$ belonging to suitable function spaces and defined such that
\begin{eqnarray}
\partial_{t}&=&N\hat{n}+X
\end{eqnarray}
with $t$ and $\hat{n}$ being time and a hypersurface orthogonal future directed timelike unit vector i.e., $\tilde{g}(\hat{n},\hat{n})=-1$, respectively. The above splitting writes the spacetime metric $\tilde{g}$ in local coordinates $\{x^{\alpha}\}_{\alpha=0}^{n}=\{t,x^{1},x^{2},....,x^{n}\}$ as 
\begin{eqnarray}
\tilde{g}&=&-N^{2}dt\otimes dt+g_{ij}(dx^{i}+X^{i}dt)\otimes(dx^{j}+X^{j}dt)
\end{eqnarray} 
and the stress-energy tensor as
\begin{eqnarray}
\mathbf{T}&=&E\mathbf{n}\otimes\mathbf{n}+\mathbf{J}\odot\mathbf{n}+\mathbf{S},
\end{eqnarray}
where $\mathbf{J}\in \mathfrak{X}(M)$, $\mathbf{S}\in S^{2}_{0}(M)$, and $A\odot B=\frac{1}{2}(A\otimes B+B\otimes A$). Here, $\mathfrak{X}(M)$ and $S^{2}_{0}(M)$ are the space of vector fields and the space of symmetric covariant 2-tensors, respectively. The choice of a spatial slice in the spacetime leads to consideration of the second fundamental form $k_{ij}$ which describes how the slice is curved in the spacetime. The trace of the second fundamental form ($tr_{g}k=\tau$) is the mean extrinsic curvature of the slice, which will play an important role in the analysis. Under such decomposition, the Einstein equations 
\begin{eqnarray}
R_{\mu\nu}-\frac{1}{2}Rg_{\mu\nu}+\Lambda g_{\mu\nu}&=&T_{\mu\nu}
\end{eqnarray}
take the form ($8\pi G=c=1$)
\begin{eqnarray}
\partial_{t}g_{ij}&=&-2Nk_{ij}+L_{X}g_{ij},\\\nonumber
\partial_{t}k_{ij}&=&-\nabla_{i}\nabla_{j}N+N\{R_{ij}+\tau k_{ij}-2k_{ik}k^{k}_{j}\\\nonumber
&&-\frac{1}{n-1}(2\Lambda-S+E)g_{ij}-S_{ij}\}+L_{X}k_{ij}
\end{eqnarray}
along with the constraints (Gauss and Codazzi equations)
\begin{eqnarray}
\label{eq:HC}
R(g)-|k|^{2}+\tau^{2}&=&2\Lambda+2E,\\
\label{eq:MC}
\nabla_{j}k^{j}_{i}-\nabla_{i}\tau&=&J_{i},
\end{eqnarray} 
where $S=g^{ij}S_{ij}$. The vanishing of the covariant divergence of the stress energy tensor i.e., $\nabla_{\nu}T^{\mu\nu}=0$ is equivalent to the continuity equation and equations of motions of the matter
\begin{eqnarray}
\frac{\partial E}{\partial t}&=&L_{X}E+NE\tau-N\nabla_{i}J^{i}-2J^{i}\nabla_{i}N+NS^{ij}k_{ij},\\
\frac{\partial J^{i}}{\partial t}&=&L_{X}J^{i}+N\tau J^{i}-\nabla_{j}(NS^{ij})+2Nk^{i}_{j}J^{j}-E\nabla^{i}N.\nonumber
\end{eqnarray}
Note that there are no evolution equations for the lapse function and the shift vector field as a consequence of gauge freedom. In a sense, the original Einstein equations (before the ADM split) consists of 6 evolution equations and 4 constraints and therefore the equations for lapse and shift should be obtained by fixing the gauge. In order to choose a time coordinate and assign uniqueness to the spatial slice, gauge fixing is required. Here we choose `constant mean extrinsic curvature' (CMC) as the temporal gauge which yields an elliptic equation for the lapse function. Later on, we will select a suitable spatial gauge. CMC gauge reads 
\begin{eqnarray}
\tau=tr_{g}k=monotonic~function~of~t~alone
\end{eqnarray}      
so $\tau$ is thus constant throughout the hypersurface and therefore can play the role of time. Using the evolution and constraint equations,  one may obtain the following equation for the lapse function
\begin{eqnarray}
\frac{\partial \tau}{\partial t}&=&\Delta_{g}N+\{|k|^{2}+\frac{S}{n-1}+\frac{n-2}{n-1}E-\frac{2\Lambda}{n-1}\}N+L_{X}\tau
\end{eqnarray}
which after implementing CMC gauge ($\partial_{i} \tau=0$) yields
\begin{eqnarray}
\label{eq:lase}
\frac{\partial \tau}{\partial t}&=&\Delta_{g}N+\{|k|^{2}+\frac{S}{n-1}+\frac{n-2}{n-1}E-\frac{2\Lambda}{n-1}\}N,\nonumber
\end{eqnarray}
where $|k|^{2}=k_{ij}k^{ij}$ and the Laplacian is defined as $\Delta_{g}=-\nabla[g]^{i}\nabla[g]_{i}$ and therefore has positive spectrum on compact connected manifolds. 

The evolution and constraint equations derived here are purely geometric and therefore one has the freedom to choose the topology of the spatial slice $M$. Different choices would lead to different cosmological models. Of course, not all topologies admit metrics that satisfy the cosmological principle. Nevertheless, we will consider a general topology to start with so that the universe may not have been isotropic and homogeneous (locally) throughout its complete course of evolution. Thurston's geometrization conjecture \cite {anderson1997scalar, cao2006complete, milnor2003towards} and its subsequent proof together with the Poincar\'e conjecture \cite {cao2006complete, morgan2007ricci, zhang2010sobolev} and its proof allow one a complete classification of compact, orientable 3-manifolds. Leaving aside the so-called `trivial' manifolds diffeomorphic to $\mathbf{S}^{3}$, the remaining possibilities consist of an infinite list of nontrivial manifolds, each of which is diffeomorphic to a finite connected sum of the following form \cite {ashtekar2015general, moncrief2019could, fischer1997hamiltonian}
\begin{eqnarray}
\label{eq:cs}
\mathbf{S}^{3}/\Gamma_{1}\#..\#\mathbf{S}^{3}/\Gamma_{k}\#(\mathbf{S}^{2}\times \mathbf{S}^{1})_{1}\#..\#(\mathbf{S}^{2}\times \mathbf{S}^{1})_{l}\#\mathbf{K}(\pi,1)_{1}\#..\#\mathbf{K}(\pi,1)_{m},
\end{eqnarray}   
where $k,l,$ and $m$ are non-negative integers satisfying $k+l+m\geq1$ and if either $k,l$ or $m$ equals to zero, then terms of that type do not occur. Here each $\Gamma_{k}$ (for $k\geq1$) is a nontrivial ($\neq I$) subgroup of SO(4) acting freely and properly discontinuously on $\mathbf{S}^{3}$ and the resulting manifold is $\mathbf{S}^{3}/\Gamma_{k}$ (see \cite {lafontaine2015introduction} and \cite {hatcher2005algebraic} for free and properly discontinuous actions). $\mathbf{S}^{2}\times \mathbf{S}^{1}$ denotes the wormhole and supports only anisotropic (even locally) homogeneous metrics and therefore is not compatible with the cosmological observations. The remaining $\mathbf{K}(\pi,1)$ manifolds are of interest to us and defined to have non-trivial fundamental group while the rest of their higher homotopy groups vanish \cite {scott1973compact, hatcher2005algebraic}. A vast set of compact hyperbolic manifolds (and therefore supporting locally homogeneous and isotropic metrics) fall under this category. In this article, we will focus on manifolds of negative Yamabe type i.e., the scalar curvature $R(g)$ of every Riemannian metric $g$ on $M$ is negative somewhere (for details about negative Yamabe manifolds, the reader is referred to \cite {ashtekar2015general, moncrief2019could, fischer1997hamiltonian}). Note that non-flat type $\mathbf{K}(\pi,1)$ manifolds are of negative Yamabe type and presence of a single negative Yamabe type manifold in the connected sum decomposition written as (\ref{eq:cs}) turns the whole manifold into one of negative Yamabe type. It is also known that every prime $\mathbf{K}(\pi,1)$ manifold is decomposable into a (finite) collection of (complete, finite volume) hyperbolic and graph manifolds (see introduction of \cite{moncrief2019could} for the detailed description of a graph manifold). This property of $\mathbf{K}(\pi,1)$ manifold will turn out to be important later in our discussion. We will, throughout our analysis, consider an `n+1' dimensional universe and only later specialize our results to the physical `3+1' universe.    

Utilizing the `CMC' condition, the momentum constraint reduces to 
\begin{eqnarray}
\label{eq:MC}
\nabla[g]_{j}k^{j}_{i}&=&J_{i},
\end{eqnarray}
the solution of which may be written as 
\begin{eqnarray}
\label{eq:2ndf}
k^{i}_{j}=K^{tri}_{j}+\frac{\tau}{n}\delta^{i}_{j}
\end{eqnarray}
with $K^{tr}$ being traceless with respect to $g$ i.e., $K^{tr}_{ij}g^{ij}=0$ (and so with respect to any metric conformal to $g$). Note that $K^{tr}$ is obtained by solving the following equation
\begin{eqnarray}
\nabla[g]_{j}K^{trj}_{i}&=&J_{i}.
\end{eqnarray}
The Hamiltonian constraint (\ref{eq:HC}) may be written using the solution of momentum constraint (\ref{eq:MC}) as follows
\begin{eqnarray}
R(g)&=&|K^{tr}|^{2}+2E-\frac{n-1}{n}(\tau^{2}-\frac{2n\Lambda}{n-1})\nonumber.
\end{eqnarray}
Since we are primarily interested in the case $\Lambda>0$, it might appear that $\tau^{2}-\frac{2n\Lambda}{n-1}$ could be negative. But then the Hamiltonian constraint would imply that if the energy density $E$ is non-negative, then $R(g)\geq0$ everywhere on $M$ which is impossible for a manifold of negative Yamabe type. Let's consider the energy condition and establish an allowed range for the constant extrinsic mean curvature $\tau$ which is now playing the role of time. The weak energy condition yields  
\begin{eqnarray}
\mathbf{T}(\mathbf{n},\mathbf{n})\geq0,\\
E\geq0
\end{eqnarray}
that is to any time-like observer the energy density is nonnegative and as such physically relevant classical matter sources are expected to satisfy this energy condition. We will only consider matter sources with point-wise nonnegative energy density throughout the spacetimes. Therefore a universe filled with matter satisfying the weak energy condition shall always have 
\begin{eqnarray}
\tau^{2}-\frac{2n\Lambda}{n-1}>0,
\end{eqnarray}
and for expanding models ($\frac{\partial \tau}{\partial t}>0$),
\begin{eqnarray}
-\infty<\tau<-\sqrt{\frac{2n\Lambda}{n-1}}.
\end{eqnarray} 
Since we are primarily interested in the asymptotic behavior of the `\textbf{expanding}' universe model (the physically relevant case), we set the range of $\tau$ to be $(-\infty, -\sqrt{\frac{2n\Lambda}{n-1}})$ once and for all. We turn our attention to the Lapse equation in an expanding universe model by setting 
\begin{eqnarray}
\label{eq:timefunction}
\frac{\partial \tau}{\partial t}&=&\frac{n}{2(n-1)}(\tau^{2}-\frac{2n\Lambda}{n-1})^{\frac{n}{2}}>0
\end{eqnarray}
whose solution $\tau=\tau(t)$ plays the role of time from now onwards. Note that (\ref{eq:timefunction}) is a valid `choice' of a time function since $\tau(t)$ is monotonic and chosen to be constant on a Cauchy hypersurface (CMC gauge). The lapse equation is explicitly written in this time co-ordinate as
\begin{eqnarray}
\label{eq:lapse}
\Delta_{g}N+(|K^{tr}|^{2}+(\frac{\tau^{2}}{n}-\frac{2\Lambda}{n-1})+\frac{1}{n-1}(S+(n-2)E))N\\\nonumber 
=\frac{n}{2(n-1)}(\tau^{2}-\frac{2n\Lambda}{n-1})^{\frac{n}{2}}.\nonumber
\end{eqnarray}
In order for this equation to have a unique positive solution, the kernel of the elliptic operator 
\begin{eqnarray}
\{\Delta_{g}+(|K^{tr}|^{2}+(\frac{\tau^{2}}{n}-\frac{2\Lambda}{n-1})+\frac{1}{n-1}(S+(n-2)E))I\}
\end{eqnarray}
defined between suitable function spaces, must be trivial. Here $I$ denotes the identity operator. If $\phi$ belongs to the kernel i.e., it satisfies 
\begin{eqnarray}
\{\Delta_{g}+(|K^{tr}|^{2}+(\frac{\tau^{2}}{n}-\frac{2\Lambda}{n-1})+\frac{1}{n-1}(S+(n-2)E))I\}\phi&=&0,\nonumber
\end{eqnarray}
then after multiplying both sides with $\phi$, integration over the closed orientable manifold $M$ and applying Stokes' theorem one gets
\begin{eqnarray}
\int_{M}\{\nabla[g]_{i}\phi\nabla[g]^{i}\phi+(|K^{tr}|^{2}+(\frac{\tau^{2}}{n}-\frac{2\Lambda}{n-1})\\\nonumber
+\frac{1}{n-1}(S+(n-2)E))\phi^{2}\}\mu_{g}=0,\nonumber
\end{eqnarray} 
where $\mu_{g}=\sqrt{\det(g)}dx^{1}\wedge dx^{2}\wedge dx^{3}\wedge.....\wedge dx^{n}$ is the volume form on $M$. Existence of a trivial kernel i.e., $\phi\equiv0$ throughout the whole of $M$, is implied by 
\begin{eqnarray}
(n-2)E+S\geq0.
\end{eqnarray}
Therefore, for a universe filled with matter sources of everywhere non-negative energy density, an additional energy condition needs to be satisfied in order to obtain a unique solution for the lapse equation. Note that matter satisfying the strong energy condition   
\begin{eqnarray}
(T_{\mu\nu}-\frac{1}{2}Tg_{\mu\nu})n^{\mu}n^{\nu}\geq0,\\\nonumber
E+S\geq0
\end{eqnarray}
falls under this category. In a sense the `weak+strong energy condition'$\subseteq \{E\geq0,S+(n-2)E\geq0, n\geq3\}$ and therefore such sources are allowed for our analysis. Of course, most physically relevant sources do satisfy the property of non-negative energy density (weak energy condition) and the attractive nature of gravity (strong energy condition). Several known sources of physical interest satisfy weak and strong energy conditions. These include for example perfect fluid and Vlasov matter (see \cite {rendall1999dynamics, lee2005einstein, andreasson2011einstein, ringstrom2013topology, fajman2019kantowski} for details about Vlasov matter). 
 
We have established the existence of a unique solution of the lapse equation provided that the matter sources in the universe satisfy a suitable energy condition. A standard maximum principle argument for the elliptic equation yields the following estimate of the lapse function 
\begin{eqnarray}
\label{eq:NES}
0&<&\frac{\frac{n}{2(n-1)}(\tau^{2}-\frac{2n\Lambda}{n-1})^{\frac{n}{2}}}{\left((\frac{\tau^{2}}{n}-\frac{2\Lambda}{n-1})+\sup(|K^{tr}|^{2}+\frac{1}{n-1}(S+(n-2)E))\right)}\leq N\leq\\\nonumber
&& \frac{n^{2}}{2(n-1)}(\tau^{2}-\frac{2n\Lambda}{n-1})^{\frac{n}{2}-1}.\nonumber
\end{eqnarray}
The important thing here is to note that the lapse function is positive for an expanding universe model. The results obtained so far will be sufficient to study the dynamical behavior in terms of a weak Lyapunov function.

\section{A weak Lyapunov function and its monotonic decay}
\cite {fischer1997hamiltonian, moncrief2019could} constructed a reduced phase space as the cotangent bundle of the higher dimensional analogue of the  Teichm\"uller space and obtained the following true Hamiltonian of the dynamics through a conformal technique 
\begin{eqnarray}
H_{reduced}&:=&\frac{2(n-1)}{n}\int_{M}\frac{\partial \tau}{\partial t}\mu_{g}.
\end{eqnarray}
In that particular case of a vacuum limit, it is indeed possible to construct such a reduced Hamiltonian in `CMC' gauge, which also acts as a weak Lyapunov function. However, as will be shown in the case of a simple matter source, the perfect fluid, the construction of a reduced Hamiltonian using a conformal diffeomorphism is restricted by certain conditions described later. However, for the purpose of the Einsteinian dynamics, we do not need to construct a reduced phase space (and associated Hamiltonian) but rather a suitable Lyapunov function and study its dynamical behavior. Let us consider the rescaled volume functional as the weak Lyapunov function and call it $L(g,k)$ 
\begin{eqnarray}
\label{eq:LF}
L(g,k)&=&\frac{2(n-1)}{n}\int_{M}\frac{\partial \tau}{\partial t}\mu_{g}.
\end{eqnarray}
We call $L(g,k)$ a weak Lyapunov function because, following the expression of $\frac{\partial\tau}{\partial t}$, it controls the $H^{1}\times L^{2}$ norm of the data (g,k) while the desired norm would be $H^{s}\times H^{s-1}, s>\frac{n}{2}+p$ for some $p\geq1$. Note that we do not, at this point, have a local existence theorem of the Cauchy problem of the Einstein system in exotic spatial topologies (negative Yamabe manifolds in this case) with arbitrary matter source satisfying the desired energy conditions. For the vacuum case, Bel-Robinson energy is used by \cite {andersson2004future}, which controls the $H^{2}\times H^{1}$ norm of the data. Recently, \cite {rodnianski2013nonlinear} proved nonlinear stability results for small data perturbations of FLRW background cosmological model, which of course deals only with special topologies ($\mathbf{S}^{3}, \mathbf{T}^{3},$ and $\mathbf{H}^{3}$). At the moment, let us focus on the time evolution of the weak Lyapunov function namely the rescaled volume functional. Due to its weak property, we would not be able to state a theorem concerning the stability (either Lyapunov or asymptotic) of the spacetime on the basis of its time evolution. Nevertheless, we will be able to obtain important physical results related to the asymptotic behavior of the spacetime (in the expanding direction) utilizing the limiting behavior of the Lyapunov function. The time evolution of this function may be obtained as  
\begin{eqnarray}
\frac{dL(g,k)}{dt}&=&\int_{M}\frac{\delta L(g,k)}{\delta g}\partial_{t}g+\int_{M}\frac{\delta L(g,k)}{\delta\nonumber  k}\partial_{t}k,\\\nonumber
&=&\frac{2(n-1)}{n}\int_{M}\left(\frac{\partial^{2}\tau}{\partial t^{2}}+\frac{1}{2}\frac{\partial \tau}{\partial t}g^{ij}\frac{\partial g_{ij}}{\partial t}\right)\mu_{g},\\\nonumber
&=&\int_{M}\tau(\tau^{2}-\frac{2n\Lambda}{n-1})^{\frac{n}{2}-1}(n\Delta_{g}+n|K^{tr}|^{2}+(\tau^{2}-\frac{2n\Lambda}{n-1})\\\nonumber &&+\frac{n}{n-1}(S+(n-2)E))N+(\tau^{2}-\frac{2n\Lambda}{n-1})^{\frac{n}{2}}\int_{M}(-N\tau+\nabla_{i}X^{i}))\mu_{g},\\\nonumber
&=&\tau(\tau^{2}-\frac{2n\Lambda}{n-1})^{\frac{n}{2}-1}\int_{M}(n|K^{tr}|^{2}+(\tau^{2}-\frac{2n\Lambda}{n-1})\\\nonumber
&&+\frac{n}{n-1}(S+(n-2)E)-(\tau^{2}-\frac{2n\Lambda}{n-1}))N\mu_{g},\\
&=&n\tau(\tau^{2}-\frac{2n\Lambda}{n-1})^{\frac{n}{2}-1}\int_{M}(|K^{tr}|^{2}+\frac{1}{n-1}(S+(n-2)E))N\mu_{g},
\end{eqnarray} 
where we have used the identity
\begin{eqnarray}
\frac{\partial^{2} \tau}{\partial t^{2}}
&=&\frac{n}{2(n-1)}\tau(\tau^{2}-\frac{2n\Lambda}{n-1})^{\frac{n}{2}-1}(n\Delta_{g}+n|K^{tr}|^{2}\\
&&+(\tau^{2}-\frac{2n\Lambda}{n-1})+\frac{n}{n-1}(S+(n-2)E))N,\nonumber
\end{eqnarray}
obtained using the lapse equation and the CMC gauge condition ($\partial_{i} \tau=0$). We have also used the Stokes' theorem to eliminate the covariant divergence terms in the integral. Along the solution curve in the expanding direction ($\frac{\partial \tau}{\partial t}>0$ and $-\infty<\tau<-\sqrt{\frac{2n\Lambda}{n-1}}$), therefore, the Lyapunov function monotonically decays i.e.,
\begin{eqnarray}
\frac{d L(g,k)}{dt}<0
\end{eqnarray}
and only attains its infimum (i.e., $\frac{dL}{dt}=0$) precisely when the following conditions are met 
\begin{eqnarray}
K^{tr}&=&0,\\
S+(n-2)E&=&0.
\end{eqnarray} 
Substitution of the first condition into the momentum constraint (\ref{eq:MC}) immediately yields 
\begin{eqnarray}
J^{i}\equiv0
\end{eqnarray}
as well as $Y\equiv0$ everywhere on $M$. In addition to satisfying $E, S+(n-2)E\geq0$, if the matter sources also satisfy the strong energy condition i.e., $(S+E)\geq0$, then for $n\geq3$ (the cases of primary interest), the second condition for the infimum of the Lyapunov function translates to
\begin{eqnarray}
S+E=0,\\
E=0
\end{eqnarray}
and therefore 
\begin{eqnarray}
S=0.
\end{eqnarray}
This result, therefore, states that the weak Lyapunov function namely the rescaled volume is monotonically decaying in the direction of cosmological expansion and approaches its nfimum only in the limit that the matter sources be `turned off' or at least become asymptotically negligible. In the limit of turned off matter, one may utilize the evolution and constraint equations by substituting $k^{TT}=\partial_{t}k^{TT}=0$ to obtain the background warped product spacetimes
\begin{eqnarray}
\label{eq:conf}
ds^{2}=-\frac{n^{2}}{(\tau^{2}-\frac{2n\Lambda}{n-1})^{2}}d\tau^{2}
+\frac{n}{(n-1)(\tau^{2}-\frac{2n\Lambda}{n-1})}\gamma_{ij}dx^{i}dx^{j},
\end{eqnarray}
with $R(\gamma)=-1$. The vital question is whether the Lyapunov function $L$ ever attains its infimum. In the expanding universe model, if the matter sources do not re-collapse to form a singularity, then the matter density falls off and in the limiting case, may be considered to be negligible. We will indeed show such asymptotic decay of the matter density in case of a perfect fluid. Observing the monotonic decay of the Lyapunov function along a solution curve, one may be tempted to conjecture the asymptotic stability of the matter filled spacetimes. However, we remind the reader again that such property only provides a weak notion of stability if the matter sources do not develop singularities. Even a vacuum spacetime might be able to go singular that is the pure gravity can collapse to form singularity before the spatial volume of the universe reaches infinity. As of now, we do not yet have a global existence theorem for even a vacuum spacetime.     
Now we specialize our results to the perfect fluid matter source. 

\section{Perfect Fluid}
The perfect fluid stress-energy tensor
\begin{eqnarray}
T^{\mu\nu}&=&(P+\rho)u^{\mu}u^{\nu}+Pg^{\mu\nu}
\end{eqnarray}
yields 
\begin{eqnarray}
E&=&T(\mathbf{n},\mathbf{n})=(P+\rho)(\tilde{g}(u,n))^{2}-P\\
&=&(P+\rho)(Nu^{0})^{2}-P\geq0
\end{eqnarray}
assuming the equation of state $P=(\gamma-1)\rho$, $1\leq\gamma\leq2$, normalization $\tilde{g}(u,u)=-1$, and a positive energy density $\rho$.
The strong energy condition requires 
\begin{eqnarray}
(nP+\rho)\geq0
\end{eqnarray}
which is clearly satisfied as well. Therefore, satisfying both of these energy conditions, a perfect fluid falls under the category for which our analysis holds true. Through a conformal technique, we will obtain the infimum of the weak Lyapunov function described in the previous section. The conformal technique may not be applied to any matter source in a unique way in general as the conformal rescaling of at least the energy density would be different for a different choice of sources. As a consequence, the resulting nonlinear elliptic equation namely the Lichnerowicz equation arising from the rescaling of Hamiltonian constraint, which will play the central role in obtaining the infimum of the Lyapunov function, may not have a unique positive solution. Nevertheless, as we will see shortly, one may obtain a suitable conformal rescaling of matter variables in the case of a perfect fluid and therefore, the subsequent analysis will follow. Using the momentum constraint, the second fundamental form may be written as follows (\ref{eq:2ndf}) 
\begin{eqnarray}
\label{eq:fundsplit}
K^{ij}=K^{trij}+\frac{\tau}{n}g^{ij},
\end{eqnarray}
where $g_{ij}K^{trij}=0$, but $\nabla_{j}K^{trij}=J^{i}$ under ths CMC condition ($\partial_{i} \tau=0$). We use the conformal transformation as described in \cite {fischer2002hamiltonian, moncrief2019could, ashtekar2015general} only replacing the momenta conjugate to the metric $g$ by the second fundamental form through a Legendre transformation $\pi=-\mu_{g}(k-(tr_{g}k)g)$. The conformal transformation reads
\begin{eqnarray}
\label{eq:conformal}
(g_{ij},K^{trij})=(\psi^{\frac{4}{n-2}}\gamma_{ij}, \psi^{-\frac{2(n+2)}{n-2}}\kappa^{trij}),
\end{eqnarray} 
where $\gamma$ and $\kappa^{tr}$ are the scale-free fields satisfying $\gamma^{ij}\kappa^{tr}_{ij}=0$ with $\gamma\in \mathcal{M}_{-1}$ and $\psi: M\to \mathbf{R}_{>0}$. Here $\mathcal{M}_{-1}$ is defined as $\mathcal{M}_{-1}=\{\gamma$ is a Riemannian metric on M $|R(\gamma)=-1\}$. 
In reality, the true dynamics assumes a metric lying in the orbit space $\mathcal{M}_{-1}/D_{0}$, $D_{0}$ being the group of diffeomorphisms (of $M$) isotopic to identity. This is a consequence of the fact that if $\gamma\in\mathcal{M}_{-1}, k, N$, and $X$ solve the Einstein equations, so do $(\phi^{-1})^{*}\gamma, (\phi^{-1})^{*}k, (\phi^{-1})^{*}N=N\circ \phi^{-1}$, and $\phi_{*}X$, where $\phi\in D_{0}$ and $^{*}$, and $_{*}$ denote the pullback and push-forward operations on the cotangent and tangent bundles of $M$, respectively. To avoid technical complexities, the calculations may be restricted to $\mathcal{M}_{-1}$ as the entities we are interested in (such as $\mu_{\gamma}=\sqrt{\det{\gamma_{ij}}}dx^{1}\wedge dx^{2}...\wedge dx^{n}$) are $D_{0}$ invariant yielding equivalence between $\mathcal{M}_{-1}$ and $\mathcal{M}_{-1}/D_{0}$ (only in this particular occasion). Note that $K^{tr}$ may be decomposed into a transverse-traceless part (with respect to $g$) and a Conformal Killing tensor part
\begin{eqnarray}
\label{eq:tracescale}
K^{trij}&=&K^{TTij}+\psi^{\frac{-2n}{n-2}}(L_{Y}g-\frac{2}{n}\nabla_{m}Y^{m}g)^{ij},
\end{eqnarray}
where $Y\in\mathfrak{X}(M)$. Under this conformal transformation, the components of the conformal Killing tensor $(L_{g}Y-\frac{2}{n}\nabla_{m}Y^{m}g)$ transforms as follows 
\begin{eqnarray}
(L_{g}Y-\frac{2}{n}\nabla_{m}Y^{m}g)^{ij}=\psi^{\frac{-4}{n-2}}(L_{\gamma}Y-\frac{2}{n}\nabla_{m}Y^{m}\gamma)^{ij}
\end{eqnarray}
yielding a consistent transformation of the transverse-traceless tensor $K^{TT}$, that is, $K^{TTij}=\psi^{-\frac{2(n+2)}{n-2}}\kappa^{TTij}$. Therefore, we may write 
\begin{eqnarray}
\label{eq:trace}
\kappa^{trij}=\kappa^{TT}+(L_{\gamma}Y-\frac{2}{n}\nabla_{m}Y^{m}\gamma)^{ij}
\end{eqnarray}

The Hamiltonian constraint, under this conformal transformation, yields the Lichnerowicz equation 
\begin{eqnarray}
\Delta_{\gamma}\psi-\frac{n-2}{4(n-1)}\psi-\frac{(n-2)}{4(n-1)}\psi^{\frac{-3n+2}{n-2}}|\kappa^{TT}+(L_{\gamma}Y-\frac{2}{n}\nabla_{m}Y^{m}\gamma)|^{2}\\\nonumber
+\frac{n-2}{4n}(\tau^{2}-\frac{2n\Lambda}{n-1})\psi^{\frac{n+2}{n-2}}
-\frac{(n-2)E[g]}{2(n-1)}\psi^{\frac{n+2}{n-2}}=0,
\end{eqnarray}
where $E[g]$ denotes the energy density without conformal scaling. Note that we can not analyze the Lichnerowicz equation without performing suitable scaling of $E[g]$ which we will precisely do in the case of a perfect fluid. But first, the momentum constraint after the rescaling reads
 \begin{eqnarray}
\nabla[\gamma]_{j}\kappa^{trij}=\psi^{\frac{2(n+2)}{n-2}}J[g]^{i},
\end{eqnarray} 
where $J[g]^{i}$ is the momentum density without conformal scaling. Now of course, if one chooses the following scaling for the momentum density (York scaling \cite {choquet2009general})
\begin{eqnarray}
J[g]^{i}=\psi^{\frac{-2(n+2)}{n-2}}J[\gamma]^{i},
\end{eqnarray}
the momentum constraint becomes decoupled from the Lichnerowicz equation i.e., 
\begin{eqnarray}
\label{eq:scalem}
\nabla[\gamma]_{j}\kappa^{trij}=J[\gamma]^{i}.
\end{eqnarray}  
Now upon substituting $\kappa^{tr}$ from equation (\ref{eq:trace}) into equation (\ref{eq:scalem}) yields an elliptic equation for $Y$
\begin{eqnarray}
-\Delta_{\gamma}Y^{i}+R[\gamma]^{i}_{m}Y^{m}+(1-\frac{2}{n})\nabla[\gamma]^{i}(\nabla[\gamma]_{m}Y^{m})&=&J[\gamma]^{i}.
\end{eqnarray}
The vector field $Y$ (not a vector density) therefore soley depends on the metric $\gamma$ and the matter source $J[\gamma]$, not on the conformal function $\psi$.
Analysis completed up to now holds for any general matter source, however, in order to obtain a rescaling of the energy density, we need to know the matter type. In the case of a perfect fluid, the momentum density is expressible in terms of the basic variables $P$ and $\rho$ which in turn yields the energy density $E$. Let us split the $n+1$-velocity vector ($u^{\mu}$) of the fluid into a hypersurface parallel component
\begin{eqnarray}
\label{eq:splitvector}
v=u+~^{n+1}g(u,\mathbf{n})\mathbf{n},
\end{eqnarray}   
and its orthogonal complement $-^{n+1}g(u,\mathbf{n})\mathbf{n}$. Here $^{n+1}g$ and $\mathbf{n}$ denote the spacetime metric and hypersurface orthogonal future directed unit vector, respectively. Note that $v_{i}=u_{i}$, but $v^{i}\neq u^{i}$. In the ADM language, the unscaled matter momentum density and energy density may be written as
\begin{eqnarray}
\label{eq:md}
J[g]^{i}&=&(P+\rho)Nu^{0}v^{i},\\
\label{eq:ed}
E[g]&=&(P+\rho)(Nu^{0})^{2}-P.
\end{eqnarray}
Now, utilizing the normalization condition $g_{\mu\nu}u^{\mu}u^{\nu}=-1$ in conjunction with the spliting (\ref{eq:splitvector}), we obtain 
\begin{eqnarray}
-(Nu^{0})^{2}+g_{ij}v^{i}v^{j}=-1.
\end{eqnarray}
Since, the right hand side of this equation is a constant (and therefore conformally invariant), each term in the left hand side has to be conformally invariant leading to the following scaling of $Nu^{0}$ and $v^{i}$
\begin{eqnarray}
(Nu^{0})_{g}&=&(Nu^{0})_{\gamma},\\
v^{i}_{g}&=&\psi^{-\frac{2}{n-2}}v^{i}_{\gamma},
\end{eqnarray}
which together with the expressions of matter energy and momentum density (\ref{eq:md}-\ref{eq:ed}) yields the scaling for $P, \rho,$ and $E[g]$
\begin{eqnarray}
(P+\rho)_{g}&=&\psi^{-\frac{2(n+1)}{n-2}}(P+\rho)_{\gamma},\\
E[g]&=&\psi^{-\frac{2(n+1)}{n-2}}E[\gamma].
\end{eqnarray}
Rescaling of the matter energy density $E[g]$ now leads to the proper Lichnerowicz equation 
\begin{eqnarray}
\Delta_{\gamma}\psi-\frac{n-2}{4(n-1)}\psi-\frac{(n-2)}{4(n-1)}\psi^{\frac{-3n+2}{n-2}}|\kappa^{TT}+(L_{Y}\gamma-\frac{2}{n\mu_{\gamma}}\nabla_{m}Y^{m}\gamma)|^{2}\\\nonumber
-\frac{(n-2)E[\gamma]}{2(n-1)}\psi^{-\frac{n}{n-2}}+\frac{n-2}{4n}(\tau^{2}-\frac{2n\Lambda}{n-1})\psi^{\frac{n+2}{n-2}}=0,\nonumber
\end{eqnarray} 
a solution of which may be obtained by the standard sub and super solution technique \cite {choquet2009general, Oxford University Press}. Note that the exponent of $\psi$ in the term $\frac{(n-2)E[\gamma]}{2(n-1)}\psi^{-\frac{n}{n-2}}$ is crucial in proving the uniqueness and existence of  solutions to the Lichnerowicz equation and therefore, the rescaling of the energy density $E[g]$ is vital. Now that we have obtained a Lichnerowicz equation for the gravity coupled fluid dynamics, we may explicitly obtain the infimum of the weak Lyapunov function. A straightforward maximum principle argument, applied to the Lichnerowicz equation, shows that the unique positive solution $\psi=\psi(\tau,\gamma, \kappa^{TT}, Y)$ satisfies 
\begin{eqnarray}
 \psi^{\frac{4}{n-2}}\geq\frac{n}{n-1}\frac{1}{(\tau^{2}-\frac{2n\Lambda}{n-1})}
\end{eqnarray}     
with equality holding everywhere on $M$ iff 
\begin{eqnarray}
\kappa^{tr}=\kappa^{TT}+\{L_{Y}\gamma-\frac{2}{n}\nabla_{m}Y^{m}\gamma\}&\equiv&0,\\
E[\gamma]&\equiv&0
\end{eqnarray}
on $M$. Therefore the asymptotic analysis becomes straightforward as the infimum of the Lyapunov function is attained precisely when the previous two conditions are met. In the limit of negligible matter (matter turned off condition), the Lyapunov function may be written in terms of the rescaled variables as follows
\begin{eqnarray}
L(\tau,\gamma,\kappa^{tr}=0)=\int_{M}(\tau^{2}-\frac{2n\Lambda}{n-1})^{n/2}
\psi^{2n/(n-2)}(\tau,\gamma,\kappa^{tr}=0)\mu_{\gamma}.
\end{eqnarray}
The infimum of the Lyapunov function over the space $\mathcal{M}_{-1}/D_{0}\times S^{0}_{2}(M)$ ($S^{0}_{2}(M)$ being the space of symmetric covariant 2-tensors) may be computed as 
\begin{eqnarray}
\inf_{\mathcal{M}_{-1}/D_{0}\times S^{0}_{2}(M)}L(\tau,\gamma,\kappa^{tr})\\\nonumber
=\inf_{\mathcal{M}_{-1}/D_{0}\times S^{0}_{2}(M)} \int_{M}(\tau^{2}-\frac{2n\Lambda}{n-1})^{n/2}\psi^{2n/(n-2)}(\tau,\gamma,\kappa^{tr})\mu_{\gamma},\\\nonumber
=(\frac{n}{n-1})^{n/2}\inf_{\mathcal{M}_{-1}/D_{0}}\int_{M}\mu_{\gamma},\\\nonumber
=(\frac{n}{n-1})^{n/2}\left\{-\sigma(M)\right\}^{n/2},
\end{eqnarray} 
where $\sigma(M)$($<0$) is a topological invariant (higher dimensional analog of the Euler characteristics of a higher genus surfaces) of the manifold $M$ (of negative Yamabe type considered here). For extensive details about the $\sigma-$constant (also known as Yamabe invariant) see \cite {petean1998computations, fischer2002hamiltonian, akutagawa2007perelman}. For the purpose here, it suffices to know that it is a topological invariant. 

The most interesting case here is the physical universe i.e., $3+1$ case. Utilizing Ricci-flow techniques, the $\sigma$ constant (and therefore the infimum of the weak Lyapunov function) of the most general compact 3-manifolds of negative Yamabe type has been computed and as such is given by 
\begin{eqnarray}
|\sigma(M)|&=&(vol_{-1}H)^{2/3},
\end{eqnarray}    
where $vol_{-1}H$ is the volume of the hyperbolic part of $M$ computed with respect to the hyperbolic metric normalized to have scalar curvature $-1$ \cite {anderson2004geometrization, anderson2005canonical}. Therefore, apart from the hyperbolic family of $\mathbf{K}(\pi,1)$ manifolds, the remaining parts of $M$ i.e., wormholes ($\mathbf{S}^{2}\times \mathbf{S}^{1}$), spherical space forms ($\mathbf{S}^{3}/\Gamma_{k}$), and the graph manifolds (non-hyperbolic part of $\mathbf{K}(\pi,1)$ manifold) do not contribute to the $\sigma$ constant. Hence, they do not contribute to the infimum of the weak Lyapunov function as well. Since the weak Lyapunov function is geometrically the rescaled volume of $M$ (\ref{eq:LF}), following its monotonic decay towards an infimum dominated only by the hyperbolic component of the spatial manifold, one is led to the natural conclusion that the Einstein flow in the presence of a suitable matter sources and positive cosmological constant drives the universe towards an asymptotic state that is volume dominated by the hyperbolic components equipped with a locally homogeneous and isotropic metric. Such results while seeming innocuous may have tremendous cosmological significance. For example, let us consider the case discussed below. A well-known result in topology is that the presence of a single negative Yamabe type component in the connected sum makes the whole manifold negative Yamabe (one may find a unique solution to the Yamabe problem) and thus the asymptotic analysis described previously holds (except stand alone flat types which are zero Yamabe type). Spaces like de-sitter ($\mathbf{S}^{3}$ with a positive cosmological constant) or the spherical space forms ($\mathbf{S}^{3}/\Gamma, \Gamma\subset$ SO(4)) or the Schwarchild-deSitter spaces ($\mathbf{S}^{2}\times \mathbf{S}^{1}$) may individually expand to infinity (proven results, see e.g., \cite {stuchlik1999some, choquet2009general}), however, while present in a connected sum with a negative Yamabe type summand, may be asymptotically volume dominated by hyperbolic parts. Let's consider the spherical space form case, for example. The spatial topology is $\mathbf{S}^{3}/\Gamma$ and the connected sum with a negative Yamabe manifold $M$ is still topologically a negative Yamabe manifold. As a consequence, the individual `expansion to infinity' property of the de-Sitter spaces (or its quotients) is sort of `killed' by the hyperbolic parts present in $M$ and therefore becomes asymptotically negligible. \cite{barrow1985} showed that spherical space forms ($\mathbf{S}^{3}/\Gamma$) and handle ($\mathbf{S}^{2}\times \mathbf{S}^{1}$) topologies re-collapse through formation of maximal hypersurfaces provided a number of conditions on matter sources and spatial geometry are satisfied (without a cosmological constant; see their theorem 3). Later \cite{barrow1986} studied the closed universe re-collapse conjecture with the same spatial topologies. This provides a rough notion that these topologies may not contribute to the asymptotic state of the universe. However, in the presence of a positive cosmological constant, spherical space forms (generalized de-Sitter spaces) may avoid formation of a maximal hypersurface and continue to expand. Our result, in this particular aspect, becomes interesting in a sense that even if $\mathbf{S}^{3}/\Gamma$ or $\mathbf{S}^{2}\times \mathbf{S}^{1}$ individually may expand forever (with a positive cosmological constant present), while present in a connected sum decomposition with hyperbolic manifolds, do not contribute to the asymptotic volume of the spatial universe due to the reason discussed previously. Now, compact quotients of hyperbolic space $\mathbf{H}^{3}$ by faithful and discrete subgroups of SO$^{+}$(3,1) provide an ample supply of compact hyperbolic manifolds equipped with locally homogeneous and isotropic metric. Therefore, the cosmological model is potentially enriched (and generalized in a certain sense) by including such rather exotic topologies capable of assuring that the cosmological principle holds.      

Following the previous analysis, two natural questions arise. Firstly, whether, as in the vacuum case (with or without a positive cosmological constant), the Lyapunov function has a physical interpretation as a Hamiltonian of the reduced dynamics that is, can one cast the problem in such a way that the reduced Hamiltonian of the dynamics naturally turns out to be the rescaled volume functional. Secondly, how does one justify the matter being asymptotically negligible? We will try to answer both of these question by invoking the variational formulation of the gravity coupled fluid dynamics developed by \cite {schutz1970perfect}, \cite {demaret1980hamiltonian}, \cite {moncrief1977hamiltonian}. A reduced Hamiltonian was constructed by \cite {demaret1980hamiltonian} in co-moving spatial gauge and a time coordinate condition, where the lapse function and shift vector field were obtained through algebraic equations. Here, however, we explicitly worked in constant mean curvature gauge and derived the monotonic decay of the weak Lyapunov function and therefore the existence of a reduced Hamiltonian in this gauge remains to be checked. Later utilizing the CMC temporal gauge together with a co-moving gauge, we will show that indeed the matter density falls off asymptotically in an expanding universe model. We will explicitly work in a `3+1' dimensional universe.     

Using Schutz's velocity potential technique \cite {schutz1970perfect}, we will attempt to construct a reduced Hamiltonian in CMC gauge. We just provide a very brief description of the Schutz's velocity potential formulation, which is necessary for our purpose. For complete details, the reader is referred to \cite {schutz1970perfect}. 

The 4-velocity of the fluid $u^{\mu}$ may be written in terms of the five velocity potentials $\phi, \alpha, \beta, \theta,$ and $S$ as
\begin{eqnarray}
u_{\mu}&=&\frac{1}{h}(\partial_{\mu}\phi+\alpha\partial_{\mu}\beta+\theta\partial_{\mu}S),
\end{eqnarray}
where $h$ is the specific enthalpy of the fluid 
\begin{eqnarray}
h=\frac{P+\rho}{n},
\end{eqnarray}
n being the baryon number density. 
From the equation of state
\begin{eqnarray}
\label{eq:eqs}
P=P(h,S),
\end{eqnarray}
we have 
\begin{eqnarray}
dP&=&ndh-nTdS,
\end{eqnarray}
T being the temperature. The equations of motions are obtained as
\begin{eqnarray}
u^{\mu}\partial_{\mu}\phi&=&-h,\\
u^{\mu}\partial_{\mu}\alpha&=&0,\\
u^{\mu}\partial_{\mu}\beta&=&0,\\
u^{\mu}\partial_{\mu}\theta&=&T,\\
u^{\mu}\partial_{\mu}S&=&0
\end{eqnarray}
along with the continuity equation 
\begin{eqnarray}
\label{eq:cont}
\nabla_{\mu}(nu^{\mu})=0,
\end{eqnarray}
where the covariant derivative is with respect to the spacetime metric (e.o.m along with continuity equation are equivalent to the vanishing of the covariant divergence of the stress energy tensor).
 The conjugate momenta corresponding to the scalar fields $q^{\eta}\equiv(\phi, \alpha, \beta, \theta, S)$ are as follows \cite {schutz1970perfect} 
\begin{eqnarray}
\label{eq:momentum1}
p^{\phi}&=&-nNu^{0},\\
p^{\alpha}&=&0,\\
p^{\theta}&=&0,\\
p^{\beta}&=&\alpha p^{\phi},\\
\label{eq:momentum2}
p^{S}&=&\theta p^{\phi}.
\end{eqnarray}
Here, note that only $\alpha, \beta,$ and $p^{\phi}$ are independent. Using Dirac's technique (for the degenerate case as the one presented here) the fluid Hamiltonian density $\mathcal{H}_{f}$ is calculated,  
which together with the gravitational Hamiltonian density provide the total Hamiltonian density of the dynamics
\begin{eqnarray}
\mathcal{H}_{total}&=&\mathcal{H}_{gravity}+\mu_{g}\mathcal{H}_{f}.
\end{eqnarray}
The lapse function $N$ and shift vector field $X$ are included in the total Hamiltonian density, which will act as Lagrange multipliers in the ADM action. Using the total Hamiltonian density, the fluid-ADM action may be written in the following  
form 
\begin{eqnarray}
 S_{ADM}=\int_{I\subset \mathbf{R}}dt\int_{\mathbf{M}}(\pi^{ij}\frac{\partial g_{ij}}{\partial t}+\sum_{\eta}p^{\eta}\dot{q}_{\eta}-\mathcal{H}_{total}) d^{n}x,\nonumber
\end{eqnarray}
where $\pi^{ij}$, the gravitational momentum conjugate to $g_{ij}$, is obtained from the second fundamental form $k^{ij}$ through the legendre transformation $\pi^{ij}=-\mu_{g}(k^{ij}-(tr_{g}k) g^{ij})$, and $q_{\eta}$ and $p^{\eta}$ are defined before. Varying the acting with respect to the Lagrange multipliers the lapse $N$ and the shift $X^{i}$, one immediately obtains the Hamiltonian and momentum constraints of the fluid coupled gravity dynamics.    

When, both of these constraints are satisfied, the term $\mathcal{H}$ totally disappears yielding the following reduced action
\begin{eqnarray}
S_{reduced}&=&\int_{I\subset \mathbf{R}}dt\int_{\mathbf{M}}(\pi^{ij}\frac{\partial g_{ij}}{\partial t}+\sum_{\eta}p^{\eta}\dot{q}_{\eta})d^{3}x.\nonumber
\end{eqnarray}
Now we peform the Hamiltonian reduction via conformal transformation. Following the conventional scaling of the metric and the momenta \cite {fischer2002hamiltonian, moncrief2019could, ashtekar2015general},
\begin{eqnarray}
(g,\pi^{TT})=(\psi^{4}\gamma,\psi^{-4}p^{TT}),
\end{eqnarray}
the conformal Killing tensor $(L_{Y}g)^{ij}-\frac{2}{3}(\nabla_{m}Y^{m})g^{ij}$ transforms as 
\begin{eqnarray}
(L_{Y}g)^{ij}-\frac{2}{3}(\nabla_{m}Y^{m})g^{ij}=\psi^{-4}((L_{Y}\gamma)^{ij}\\\nonumber
-\frac{2}{3}(\nabla[\gamma]_{m}Y^{m})\gamma^{ij}).
\end{eqnarray}
Using the Legendre transformation $\pi^{ij}=-\mu_{g}(k^{ij}-(tr_{g}k) g^{ij})$ and the equations (\ref{eq:fundsplit}-\ref{eq:tracescale}), the gravitational momenta $\pi^{ij}$ may be written as follows
\begin{eqnarray}
\pi^{ij}&=&\pi^{TTij}+\frac{tr_{g}\pi}{3}g^{ij}-\psi^{-6}\mu_{g}[(L_{Y}g)^{ij}-\frac{2}{3}(\nabla_{m}Y^{m})g^{ij}],
\end{eqnarray}
where $\pi^{TT}$ is transverse-traceless with respect to $g$ i.e., $\pi^{TTij}g_{ij}=0=\nabla[g]_{j}\pi^{TTij}$.

Using these conformal transformations, the reduced action becomes 
\begin{eqnarray}
S_{reduced}
&=&\int_{I\subset \mathbf{R}}dt\int_{M}(p^{TTij}\frac{\partial \gamma_{ij}}{\partial t}-\frac{4\mu_{g}}{3}\frac{\partial \tau}{\partial t}\\\nonumber
&&-((L_{Y}\gamma)^{ij}-\frac{2}{3}\nabla[\gamma]_{m}Y^{m}\gamma^{ij})\frac{\partial \gamma_{ij}}{\partial t}\mu_{\gamma}\\\nonumber
&&+\sum_{\eta}p^{\eta}\dot{q}_{\eta})d^{3}x,
\end{eqnarray}
where we have left the fluid variables unscaled intentionally for reasons which will be discussed now. Note that even though Schutz's formalism is only true for `3+1' spacetimes, the result we are about to state holds for any higher dimension irrespective of the fluid source. That is any non-vanishing momentum flux density $J^{i}$ leads to the following problem while trying to construct an unconstrained Hamiltonian. Following the conformally scaled reduced action, one immediately observes that in order to obtain a reduced Hamiltonian, one must reduce the action to the form similar to $\int_{I\times M}dtd^{3}x(\pi^{TT}\partial_{t}\gamma+\sum_{\eta}p^{\eta}\dot{q}_{\eta}-\mathcal{H}_{reduced})$ and therefore, one need to somehow make the term $\int_{M}((L_{Y}\gamma)^{ij}-\frac{2}{n}\nabla[\gamma]_{m}Y^{m}\gamma^{ij})\frac{\partial \gamma_{ij}}{\partial t}\mu_{\gamma}$ disappear. In order to do so, we must work in a slice of the $D_{0}$ action on $\mathcal{M}_{-1}$ i.e., $\frac{\partial \gamma_{ij}}{\partial t}=\frac{\partial \gamma_{ij}}{\partial r^{a}}\dot{r}=l^{TT}_{ij}\dot{r},$ for some $\gamma-$transverse-traceless covariant 2-tensor $l$ and co-ordinates $\{r^{a}\}$ on some local chart of $\mathcal{M}_{-1}/D_{0}$. However, for dimension $n>2$, such procedure is restricted as a consequence of the absence of a globally integrable distribution i.e., a slice of the $D_{0}$ action is not a submanifold of $\mathcal{M}_{-1}$ in general (and therefore is not integrable). For 2 dimensions, the Teichm\"uller space has a nice global manifold structure (as a submanifold of $\mathcal{M}_{-1}$) and thus is globally integrable. Of course, in the absence of a matter source (with or without cosmological constant), the vector field Y, the generator of the diffeomorphism of $M$ vanishes due to the momentum constraint and the weak Lyapunov function $\frac{4}{3}\int_{M}\frac{\partial \tau}{\partial t}\mu_{g}$ becomes a reduced Hamiltonian.\\
Even though we could not obtain a reduced Hamiltonian in CMC gauge, we may show using CMC time gauge and a co-moving gauge, that the matter sources (perfect fluid), in an expanding universe, becomes asymptotically negligible. Let us write the time and spatial slice gauge explicitly first. In `3+1' CMC gauge, a time coordinate is chosen as (consistent throughout the study)
 \begin{eqnarray}
 \frac{\partial \tau}{\partial t}&=&\frac{3}{4}(\tau^{2}-3\Lambda)^{\frac{3}{2}}
 \end{eqnarray}  
along with the co-moving spatial gauge condition 
\begin{eqnarray}
u^{i}&=&0,
\end{eqnarray}
which through the normalization condition $g_{\mu\nu}u^{\mu}u^{\nu}=-1$ yields
\begin{eqnarray}
\label{eq:algb}
u^{0}&=&\frac{1}{\sqrt{N^{2}-X_{i}X^{i}}}
\end{eqnarray}
with $N^{2}>X_{i}X^{i}$. 
Noting that both $E$ and $S$ are nonnegative for perfect fluids, an estimate of $N$ follows immediately from (\ref{eq:NES})
\begin{eqnarray}
0&<&\frac{\frac{9}{4}(\tau^{2}-3\Lambda)^{\frac{1}{2}}}{\left(1+\frac{1}{\tau^{2}-3\Lambda}\sup(|K^{tr}|^{2}+\frac{1}{2}(S+2E))\right)}\leq N\leq\frac{9}{4}(\tau^{2}-3\Lambda)^{\frac{1}{2}}.\nonumber
\end{eqnarray}
Since the condition $N^{2}>X_{i}X^{i}$ has to be satisfied point-wise through the spacetime, $X_{i}X^{i}$ may be written in terms of $N^{2}$ as $CN^{2}$ for a suitable $1>C>0$. Finally, the continuity equation (\ref{eq:cont}) may be written in the chosen gauge as
\begin{eqnarray}
\frac{\partial}{\partial t}(\sqrt{N\mu_{g}}nu^{0})&=&0,
\end{eqnarray} 
which upon substituting $u^{0}$ and $X_{i}X^{i}$ yields 
\begin{eqnarray}
n\approx C_{1}\sqrt{\frac{N}{\mu_{g}}},
\end{eqnarray}
where $C_{1}>0$ is a suitable constant. In an expanding universe (if initially expanding, it will only continue to expand, because, the formation of a maximal hypersurface and subsequent re-collapse are strictly prohibited; $\tau$ can never vanish by virtue of the Hamiltonian constraint)  , the spatial volume form $\mu_{g}$ satisfies the estimate $\mu_{g}\approx (\tau^{2}-3\Lambda)^{-2\alpha},$ for some $\alpha>0$ (may be obtained by time differentiating $\int_{M}\mu_{g}$ in conjunction with the use of field equations and estimate of lapse function), which together with the estimate of $N$ yields the decay of the baryon density 
 \begin{eqnarray}
 n\approx C_{2}(\tau^{2}-3\Lambda)^{\frac{1}{4}+\alpha},
 \end{eqnarray}
for some $C_{2}>0$. In the asymptotic limit $\tau\to-\sqrt{3\Lambda}$, the baryon density becomes negligible if the matter source does not already collapse to form a singularity. Therefore, the previous asymptotic analysis holds true and the infimum of the Lyapunov function is attained in this limit allowing the persistence of topological consequences discussed previously. However, an important thing to note here is that the analysis presented here does not address the global existence by any means (despite the fact that the Lyapunov function provides a weak notion). Even in the absence of matter sources, pure gravity could `blow up' before the volume tends to infinity i.e., gravitational singularities could prevent global existence. Such global existence is addressed by \cite {andersson2011einstein} in the special case of small data limit and vacuum spacetimes (without cosmological constant) of arbitrary dimension. In case of vacuum spacetimes with a positive cosmological constant, proof of the global existence of the sufficiently small perturbations about background conformal spacetimes (eq. \ref{eq:conf}; through studying the asymptotic stability of such spacetimes) is under preparation by the author \cite {puskar2019}. \cite {fajman2015stable} studied the global existence of small data perturbations in case of vacuum spacetimes with a cosmological constant by establishing the Lyapunov stability of the background solutions. In the presence of matter sources, global existence, even in the small data limit, is an open problem. 

Developing a cosmological model is of fundamental importance in general relativity. The present so-called FLRW models are developed based on astronomical observations of local homogeneity and isotropy (since the observations are only limited to a possibly quite small fraction of the universe) that is the procedure resembles `model fitting' in certain sense. Extrapolating the local observations to a global scale thereby allows only special simple topologies and leads to the reduction of Einstein equations to ordinary differential equation for one unknown namely the scale parameter. This is, in a sense, against the whole spirit of general relativity where one should include more general topologies and study the Einsteinian dynamics in a rigorous way. The resulting cosmological model should then be compared locally with the observations. For example, for a compact spatial topology, within the injectivity radius around a point, space seems locally flat despite the possibility that the manifold may originally be globally exotic (with strictly nonvanishing curvature). In addition, manifolds may also admit locally homogeneous and isotropic metrics with no such global extension. In a sense, the role of spatial topologies in the dynamics of general relativity is somewhat underestimated and we precisely address possibly a small fraction of such roles here. We show by invoking a weak Lyapunov function that while the initially expanding universe in the presence of suitable matter sources (satisfying suitable energy conditions) and a positive cosmological constant, may not be isotropic and homogeneous (not even locally), it nevertheless will evolve to support a homogeneous and isotropic metric if the evolution continues long enough without going singular. Therefore, one does not in general need to start the evolution with an isotropic and homogeneous metric at all. Einsteinian evolution automatically leads to the spatially homogeneous and isotropic universe in the asymptotic limit where the baryon number density becomes negligible. Our analysis of course only addresses the issue of asymptotic negligibility of the matter in the case of a rather special perfect fluid source through invoking Schutz's velocity potential formalism in a constant mean curvature co-moving gauge. In the limiting case, as shown previously, the asymptotic vanishing of the matter density simplifies the associated conformal equation (Lichnerowicz equation) leading us to the main result. Of course, during the course of evolution, matter or gravitational radiation may collapse to form a singularity and a rough notion of such singularity formation (or at least geodesic incompleteness via formation of caustics) is stated by the singularity theorems for matter sources satisfying the strong energy condition. In our case while a positive cosmological constant is included, even the formation of caustics is excluded in an expanding universe. In order to establish an absolute claim that `true' singularities are avoided, one must investigate the dynamics of Einstein equations in the presence of exotic topologies numerically (which is currently only limited to very special manifolds with several symmetries). Nevertheless, it is realized through our analysis that the universe may be locally isotropic and homogeneous, but does not have to be globally so. Such topological non-triviality demonstrated here, may be only the beginning of a bigger picture and may potentially be able to answer the fundamental questions related to issues such as the flatness problem, horizon problem or even have consequences for the \textbf{Cosmic Censorship} conjecture. 

Another interesting fact revealed through our analysis is that while the associated Lyapunov function has the physical meaning of being a reduced Hamiltonian (`true Hamiltonian of the dynamics) only in case of a vacuum universe (with or without a positive cosmological constant), and therefore may be used for the purpose of quantization, matter sources (even the special perfect fluid case) rule out such significance in CMC gauge. Nevertheless, we are able to draw general conclusions about the asymptotic behavior of the spacetimes using the Lyapunov function alone. However, the weak nature (in terms of norm control) as discussed previously, does not allow a global existence theorem and for the time being we are only limited to the asymptotic results. Nevertheless, our analysis opens up several interesting questions which as described before may potentially address several existing problems. Are the Cauchy hypersurfaces always asymptotically volume-dominated by their hyperbolic components with the rescaled metrics on these components asymptotically approaching homogeneity and isotropy? Do wormholes ($\mathbf{S}^{2}\times\textbf{S}^{1}$) , spherical space forms ($\mathbf{S}^{3}/\Gamma, \Gamma\in$SO(4)), and graph manifolds always asymptotically pinch off from the universe? How is the fundamental question of `Cosmic Censorship' influenced by answers to these questions? what happens if one includes matter sources not satisfying either or both of the energy conditions mentioned? Upon further development of Einsteinian dynamics with matter sources, these questions may be answered yielding a self consistent cosmological model.   

\section{Acknowledgement}
P.M would like to thank Prof. Vincent Moncrief for numerous useful discussions related to this project and for his help improving the manuscript. This work was supported by Yale university.

\end{document}